\documentclass[letterpaper]{article} 
\usepackage{aaai25}  
\usepackage{times}  
\usepackage{helvet}  
\usepackage{courier}  
\usepackage[hyphens]{url}  
\usepackage{graphicx} 
\usepackage{natbib}  
\usepackage{caption} 
\usepackage{algorithm}
\usepackage{algorithmic}
\usepackage{graphicx}
\usepackage{textcomp}
\usepackage{xcolor}
\usepackage{amsmath} 
\usepackage{multirow}
\usepackage{makecell}
\usepackage{newfloat}
\usepackage{listings}
\DeclareCaptionStyle{ruled}{labelfont=normalfont,labelsep=colon,strut=off} 
\floatstyle{ruled}
\newfloat{listing}{tb}{lst}{}
\floatname{listing}{Listing}
\title{A Novel Breast Ultrasound Image Augmentation Method Using Advanced Neural Style Transfer: An Efficient and Explainable Approach}
\author{
    Lipismita Panigrahi,
    Prianka Rani Saha,
    Jurdana Masuma Iqrah,
    Sushil Prasad
}
\affiliations{
    The University of Texas at San Antonio\\

    lipismita.panigrahi@utsa.edu\\priankarani.saha@my.utsa.edu,\\ jurdanamasuma.iqrah@utsa.edu, \\ sushil.prasad@utsa.edu

}

\usepackage{bibentry}

\begin{document}

\maketitle

\begin{abstract}
Clinical diagnosis of breast malignancy (BM) is a challenging problem in the recent era. In particular, Deep learning (DL) models have continued to offer important solutions for early BM diagnosis but their performance experiences overfitting due to the limited volume of breast ultrasound (BUS) image data. Further, large BUS datasets are difficult to manage due to privacy and legal concerns. Hence, image augmentation is a necessary and challenging step to improve the performance of the DL models. However, the current DL-based augmentation models are inadequate and operate as a black box resulting lack of information and justifications about their suitability and efficacy. Additionally, pre and post-augmentation need high-performance computational resources and time to produce the augmented image and evaluate the model performance. Thus, this study aims to develop a novel efficient augmentation approach for BUS images with advanced neural style transfer (NST) and Explainable AI (XAI) harnessing GPU-based parallel infrastructure. We scale and distribute the training of the augmentation model across 8 GPUs using the Horovod framework on a DGX cluster, achieving a 5.09 speedup while maintaining the model’s accuracy. The proposed model is evaluated on 800 (348 benign and 452 malignant) BUS images and its performance is analyzed with other progressive techniques, using different quantitative analyses. The result indicates that the proposed approach can successfully augment the BUS images with 92.47\% accuracy.
\end{abstract}


\section{Introduction}
\label{Introduction}
The rising prominence of deep learning (DL) based techniques to be explainable is becoming more paramount, especially in domains where critical decisions must be made quickly, like medical image analysis. However, among all the cancers, breast malignancy (BM) is the most prevalent disease and is the fifth leading cause of cancer-related death among women worldwide \cite{atrey2024multimodal}. As per the World Health Organization (WHO), in 2020, approximately 2.3 million new BM were reported with 6,85,000 deaths due to this disease. Further, according to the American Cancer Society in the United States, there will be 310,720 new cases were reported for BM and 56,500 cases of Ductal Carcinoma In Situ (DCIS) will be identified as of the year 2024 
. Therefore, early diagnosis of BM may decrease the disease's effects, reduce the chance of infection, and improve medical treatment. In place of biopsy and physical examination, medical imaging (breast magnetic resonance imaging (MRI), mammogram, and, ultrasound (US), etc.) is an essential component in the timely detection of BM \cite{atrey2023mammography, panigrahi2024mbccf, panigrahi2022evaluation}. Although biopsy is the current gold standard procedure for BM diagnosis, due to the lack of labs in remote locations and, the painful, inconvenient, and time-consuming process, US imaging is an essential alternative in comparison to MRI and mammography for diagnosing BM. Also, when comparing MRI and mammography, the US has a number of advantages such as enhanced sensitivity, radiation-free imaging, cheap cost, and convenient accessibility \cite{panigrahi2019ultrasound, zhu2024dbl, panigrahi2018hybrid}. However, due to significant drawbacks of US images like poor contrast and resolution, indistinct margins caused by noise such as speckles, faint surrounding tissue, and acoustic shadowing, the breast US (BUS) image-based diagnostic method is extremely operator-reliant.

To overcome these issues computer-aided detection (CAD) system using breast US images is introduced by the researcher \cite{zhu2024dbl, panigrahi2018hybrid, rahmani2024automatic, wang2024mf}. In order to reduce operator dependency, inter and intra-observer variability and offer an alternate method for the automated and early BM diagnosis, the current CAD systems made use of machine learning (ML) and DL techniques \cite{panigrahi2024mbccf}. Further, the performance of the DL models significantly depends on training data set size and requires human annotation by radiologists. The latter is a time-consuming task. Therefore, publicly accessible BUS image data sets are small resulting in over-fitting and unable to provide generalized output. Moreover, large BUS datasets are difficult to manage because of privacy and legal concerns. Hence, image augmentation is a necessary and challenging step to enhance the performance of the DL models and regularize the overfitting \cite{xu2024automatic}. Though existing DL models have achieved some success in this direction, their performance and suitability for augmentation in BUS images are still unsatisfactory \cite{oza2022image}. Further, most researchers use the traditional DL models as black-box models for image augmentation \cite{sarp2023xai}. These cutting-edge DL models lack of the information and justifications to support the radiologists for better decisions and interpretations of BM. This opportunity is made possible by explainable AI (XAI), which converts DL-based black-box models into more transparent and understandable gray-box models. Additionally, pre and post-augmentation need high-performance resources for computation and time to produce the augmented image and evaluate the model performance. Thus, integrated data-parallel distributed training on high-performance computing systems built with GPUs becomes essential \cite{masuma2024parallel, koo2023automated}.

This motivates us to develop a novel augmentation approach for BUS images with advanced Neural Style Transfer (NST) and XAI harnessing GPU-based parallel infrastructure. The proposed augmentation approach is three-fold such as 1) Initially we introduced a novel style loss function by combining the style loss from demystifying NST (DNST) \cite{li2017demystifying} and $mr^{2}$NST \cite{wang2020mr} model, 2) further, A XAI based layer-wise relevance propagation (LRP) method is applied to the content loss function of the proposed NST model for explaining the importance of the features. This model is a post-hoc XAI model that computes the relevance score to indicate the significance of a feature in the input image, 3) To improve the efficiency of the augmentation model a distributed deep learning model training using Horovod framework on a DGX cluster that executes on 8 GPUs with a nearly linear speedup of 5.09 is used. Finally, the efficacy of the presented model is determined through quantitative analysis.
\subsection*{Contributions and paper outline }
\label{sec:Contributions and outline of the paper}

After identifying the shortcomings in current research (covered in \ref{sec:Origin of the Problem}), the following are the proposed approaches this study presents to address these issues:

\begin{itemize}
    \item A novel NST-based augmentation model is proposed by combining the demystifying NST (DNST) \cite{li2017demystifying} and $mr^{2}$NST \cite{wang2020mr} model. In this model the two loss components utilize a pre-trained ResNet50 model to extract semantic features from images as well as to resolve overfitting and gradient degradation issues.
   
    \item An XAI-based LRP method is integrated with the content loss function of the proposed augmentation model and a heatmap is generated to explain the importance of the extracted features for the model's decision-making process. By elucidating key features, this model aims to improve the interpretability and performance of the deep learning model, facilitating more generalized outputs. 
        
     \item To improve the efficiency of the augmentation model, a distributed deep learning model training leveraging Horovod on a DGX is employed that executes on 8 GPUs, yielding 5-fold speed up compared to the single-GPU execution time.

    \item To classify pre and post-augmented BUS images, we utilized a fully fine-tuned ResNet50 model to improve the accuracy and to support the radiologists for better decisions and interpretations of breast malignancy. The ResNet50 trained with augmented data yields 92.47\% accuracy and it increased by 37.26\% from the pre-augmented BUS images.

    \item To demonstrate the efficacy of the proposed model, we compared our model with cutting-edge methods using benchmark evaluation metrics.
    
\end{itemize}

The remaining paper is organized as follows. Section \ref{sec:RelatedWork} presents the related work and theoretical background on the augmentation model. Section \ref{sec:Material and Methods}, covered the materials and techniques utilized in the suggested methodology. The experimental results and discussions are described in Section \ref{sec:Experimental results and discussion}, and finally, this study is wrapped up in Section \ref{sec:Conclusion}.

\section{Related work}
\label{sec:RelatedWork}
In recent years, several studies have been published highlighting the image augmentation for medical imaging using the DL models. In Oza, P. \cite{oza2022image}, the author summarizes different types of image augmentation techniques for medical images. Generally, conventional image augmentation techniques utilize fundamental procedures to produce training images such as geometric transformation (e.g. flipping, rotation, translation, and scaling), pixel level augmentation (e.g., noise addition, sharpening), pseudo-color augmentation, random erasing, kernel filters \cite{xu2024automatic}. However, these techniques choose a transformation sequence at random for each image or depend on human professionals who have previous experience with the dataset to create a transformation sequence that will be used in training, which leads to training samples that are not sufficiently diverse. Thus, the researcher has proposed some advanced augmentation techniques such as generative adversarial networks (GAN) \cite{al2019deep}, neural style transfer (NST) \cite{georgievski2019image}, SaliencyMix \cite{uddin2020saliencymix}, Random Erasing, and Keep Augment \cite{zhong2020random}. Though all of these techniques increase the model's accuracy, they are all limited by the incapacity of feedback changes in the model's accuracy to modify the image augmentation network. Therefore, figuring out which data augmentation technique best boosts the model's performance is challenging.

To overcome this problem Cubuk \textit{et. al.} \cite{cubuk2019autoaugment} proposed an automated search for improved data augmentation policies. This study maximizes the classifier's performance by automatically searching for a sequence of transformations. Though, this method lowers the classification error rate, simultaneously computationally intensive and time-consuming due to repeatedly training the sub-model enough with various augmentation rules. Nevertheless, each image has a different ideal set of modifications, thus few augmented images might not be helpful or even detrimental for training models \cite{xu2024automatic}. Further, to overcome this issue Xu \textit{et. al.} \cite{xu2024automatic} proposed an adaptive sequence-length-based deep reinforcement learning (ASDRL) model for autonomous data augmentation in medical image analysis. This study addresses issues like partially augmented images being insufficient and being ineffective by the automatic stopping mechanism (ASM) and the reward function with dual restrictions.

Another work by Georgievski \textit{et. al.} \cite{georgievski2019image} demonstrated image augmentation methods with NST. Later on, Ramadan \textit{et. al.} \cite{ramadan2020using} enhanced Convolutional Neural Network (CNN)'s effectiveness in BM detection in Mammograms images by combining data augmentation with a cheat sheet that included common features derived from region of interest (ROI), resulting in 12.2\% and 2.2\% increase in accuracy and precision respectively. However, there may be domain gaps between medical images collected from different modalities and suppliers and their highly distinctive visual styles, which might hamper DL models. Wang \textit{et. al.} \cite{wang2020mr} proposed a multi-resolution and multi-reference NST ($mr^{2}$NST) network to address style variation in mammograms. Later on, Oza \textit{et. al.} \cite{oza2022image} summarize the image augmentation methods for mammogram evaluation. In addition to the literature already mentioned, a few earlier works summarized the basic and advanced augmentation approaches in Table \ref{table1}. 

However, in literature various research works on high-performance computing systems (HPC) with GPUs are discussed to train the DL model such as this study \cite {sergeev2018horovod}  introduced Horovod, an open-source library to train the DL models with large amounts of computation, that improves on both obstructions to scaling. Later on, Zhang, R. et. al. \cite{zhang2020super} proposed a distributed deep learning model using HPC with GPUs to speed up the learning of the unexplored low to high-resolution mapping of large volumes of Sentinel-2 image data. Further, this paper \cite{masuma2024parallel} used the Horovod-based distributed deep learning model to train a U-Net model, achieving a linear speedup of 7.21x using 8 GPUs.

\begin{table}
\centering
\begin{tabular}{p{0.2cm}p{1.1cm}p{1cm}p{2.4cm}p{2.4cm}}
\hline
SL. NO.	& Augme-ntation  approaches & Label Preserving 
& Pros	& Cons\\ \hline
1	&Geometric Transformation &No&
	Effective solution for the training data having positional bias and simple to implement.& Extra memory, computational cost is high for Transformation, computational time is more for training, and manual observation.\\

2	&Noise Injection &Yes&Enables more robust learning of the model&It's hard to decide the noise amount to be added.	\\
	
3&	Kernel Filters &Yes&Good for producing blurry and sharpened images&
	Comparable to the Constitutional Neural Network (CNN) approach.\\
	
4&	Mixing Images &No	&-&	Unfit for use in medical imagery\\
	
5&	Random Erasing &Not always	&comparable to regularization for dropouts. 
Developed to overcome occlusion-related image identification difficulties, A promising method to ensure that a network looks at the whole image rather than just a portion of it.	&Certain manual interventions can be required based on the application and dataset.\\

6&	Adversarial Training &
	Yes&
	Assist in more effectively illuminating flimsy decision boundaries than conventional categorization metrics	&Less investigation\\

7	&Generative Adversarial Network &	Yes& GANs produce data that resembles the source data.&	Augmented data are more difficult to train.\\
	
8&	Neural Style Transfer &	-&	Increases the simulated datasets capacity for generalization& Augmentation needs high-performance computational resources and time and the effort involved in choosing a style.\\ \hline	
\end{tabular}
\caption{ Basic and
 advanced augmentation approaches \cite{oza2022image}.}
 \label{table1}  
\end{table}

\subsection{Theoretical background}
\label{sec:Theoretical background}
\subsubsection{Transfer Learning based augmentation model}
\label{sec:Trasfer Learning based Augmentation Model}
 Gatys \textit{et. al.} \cite{gatys2016image} first introduced the NST. It aims to generate a stylized image $\hat{I}$ using two input images: a content image $I_{c}$ and a style reference image $I_{s}$. Then it proceeds to learn features from the content image through feature representations of $F_{l}(I_{c})$ and $F_{l}(I_{s})$ in layer \textit{l} of an NST model. Each layer of feature maps, represented by $F_{l}(I)$, where $F_{l} \in R ^{{M_l}(I)}\times N_l$ shows different aspects of style. Here, $N_l$ denotes the number of feature maps in layer \textit{l}, and $M_l (I) = H_l \times W_l (I)$ representing their spatial dimensions. The output image $\hat{I}$ in NST is generated by minimizing a loss function shown in Eq. \ref{eq1}.
\begin{equation}
\label{eq1}
L_{total}=\alpha L_c (I_c)+\beta L_s (I_s)
\end{equation}
where $\alpha$ and $\beta$ represent as the weights of the style and content losses respectively. These two loss components utilize a pre-trained VGG19 model to extract semantic features from images. The $L_c$ contrasts the feature maps of the $\hat{I}$ and $I_{c}$ defined in Eq. \ref{eq2} and $L_s$ contrasts the sum of summary statistics defined in Eq. \ref{eq3}.
\begin{equation}
\label{eq2}
L_{c}=\dfrac{1}{2}\sum_{i=1}^{N_{l}}\sum_{j=1}^{M_{l}}{\Biggl( F_{l_{c}}(\hat{I})-F_{l_{c}}(I_c)\Biggl)}^{2}_{ij}
\end{equation}

\begin{equation}
\label{eq3}
L_{s}=\sum w_{l}E_{l}
\end{equation}
where, in the layer \textit{l}, $w_l$ is the weight of the loss and $E_{l}$ the squared error between the feature correlations provided by Gram matrices defined in Eq. \ref{eq4}. The Gram matrices $G_{l}(\hat{I})$ is the inner product between the feature map $\hat{I}$ in layer \textit{l} shown in Eq. \ref{eq5}.
\begin{equation}
\label{eq4}
E_{l}={\dfrac{1}{4{N_{l}^2}M_{l}^2}}\sum_{i=1}^{N_{l}}\sum_{j=1}^{N_{l}}{\Biggl( G_{l}(\hat{I})-G_{l}(I_s)\Biggl)}^{2}_{ij}
\end{equation}
\begin{equation}
\label{eq5}
G_{l}({I})=\sum_{k=1}^{M_{l}}{ F_{l}(I)^TF_{l}(I)}
\end{equation}
However, according to \cite{li2017demystifying}, the representation of style by the Gram matrix is unclear in the cutting-edge NST model. Thus, \cite{li2017demystifying} proposed a modified the $L_{style}$ function by minimizing the Maximum Mean Discrepancy (MMD) (for detailed description about MMD please see Eq. 1,2 and 8 in \cite{li2017demystifying} ) using the second order polynomial kernel described in Eq. \ref{eq9}.

\begin{equation}
\begin{split}
\label{eq9}
L_{MMDstyle}=\dfrac{1}{4{N_{l}^2}M_{l}^2}\sum_{k_{1}=1}^{M_{l}}\sum_{k_{2}=1}^{M_{l}} \biggl(k(f_{l_{\cdot k_1}}, f_{l_{\cdot k_2}})\\
+k(f_{l}(I_s)_{\cdot k_1}, f_{l}(I_s)_{\cdot k_2})\biggl)\\
=\dfrac{1}{4N_l^2}MMD^2[F_l,F_l(I_s)]
\end{split}
\end{equation}

Though the NST model proposed in \cite{gatys2016image} is a benchmarking method for image augmentation, it suffers from various limitations:
\begin{itemize}
    \item support images with a resolution less than 1000 × 1000. Due to the greater dimensions of the medical images like 2000 × 2000, the degree of mild anomalies could be affected.
  \item The style reference image chosen manually for model input
\end{itemize}
 To address the first shortcoming \cite{wang2020mr} used a multi-resolution model and for the second issue \cite{wang2020mr} study proposed multi-reference NST (mrNST) model shown in Eq. \ref{eq6} and the necessary condition are shown in Eq. \ref{eq7} and \ref{eq8}. 

\begin{equation}
\label{eq6}
L_{multi-ref}=\sum w_lE_L
\end{equation}
where, 
\begin{equation}
\label{eq7}
E_{l}={\dfrac{1}{4{N_{l}^2}M_{l}^2}}\sum_{i=1}^{N_{l}}\sum_{j=1}^{N_{l}}{\Biggl( G_{l}(\hat{I})-G_{l}\Biggl)}^{2}_{ij}
\end{equation}

\begin{equation}
\label{eq8}
G_{l}= H(M(F_{l}(I_{s_1}),F_{l}(I_{s_2}),...F_{l}(I_{s_n})),\overline{h})
\end{equation}

The function \textit{M()} performs an element-wise max operation on a set of feature maps $F(I_{S_n})$ corresponding to the $n^{th}$ reference image at the $l^{th}$ layer. These feature maps have dimensions of $nN_l \times H_l \times W_l$. The output of this operation is a square matrix $G_l$ with dimensions $N \times N$. Each element $g_{ij}$ is computed as the maximum value obtained from the corresponding elements across all feature maps. \textit{H} is a histogram specification function and $\overline{h}$ is the density histogram. 

\subsubsection{Layer-wise relevance propagation (LRP)}
\label{sec:Layer-wise relevance propagation (LRP)}
LRP is a backward propagation method implemented for delivering explanations and performing exceptionally well across multiple benchmarks \cite{montavon2018methods}. The LRP model is a post-hoc XAI model. In \cite{binder2016layer}, the LRP method is extended to Neural Network (NN) that include local renormalization layers and a common product-type non-linearity found in DNN. Further, \cite{montavon2018methods} used the LRP method to offer theories, suggestions, and tips for utilizing it on actual data as effectively as possible. This study proves that gradient-based methods like LRP are more significant in detecting negatively relevant areas as compared with guided backpropagation.

Let $R_j^{l+1}$ denote the relevance of neuron \textit{j} at a network layer \textit{l+1} with respect to the prediction \textit{f(x)}. Then, decompose $R_j^{l+1}$ into message $R_{i\leftarrow j}^{(l,l+1)}$ sent to those neurons \textit{i} at the preceding layer \textit{l} to provide inputs to \textit{j} which presented in Eq. \ref{eq10}.

\begin{equation}
\label{eq10}
R_{j}^{(l+1)}=\sum_{i\in (l)}R_{i\leftarrow j}^{(l,l+1)}
\end{equation}

Similarly, the relevance of a neuron \textit{i} at lower layer \textit{l} accumulates all messages from the neurons in the higher layer \textit{l+1}, described in Eq. \ref{eq11}.

\begin{equation}
\label{eq11}
R_{i}^{(l)}=\sum_{j\in (l+1)}R_{i\leftarrow j}^{(l,l+1)}
\end{equation}

Based Eq. \ref{eq131} the combination of the Eqs. \ref{eq10} and \ref{eq11} guarantees a conservation property of importance between layers. So, the chain of equalities is presented in Eq. \ref{eq12}

\begin{equation}
\label{eq131}
\sum_{i}R_{i}^{(l)}=\sum_{i}\sum_{j}R_{i\leftarrow j}^{(l,l+1)}=\sum_{j}\sum_{i}R_{i\leftarrow j}^{(l,l+1)}=\sum_{j}R_{j}^{(l+1)} 
\end{equation}

\begin{equation}
\label{eq12}
\sum_{k=1}^{d}R_{k}=...=\sum_{i}R_{i}^{l}=\sum_{j}R_{j}^{(l+1)}=...=f(x)
\end{equation}

\subsection{Origin of the problems}
\label{sec:Origin of the Problem}
Though NST, $mr^{2}$NST, and demystifying NST are the benchmarking methods for image augmentation, even so, they suffer from the following downsides.

\begin{itemize}
     \item In $mr^{2}$NST, the VGG19 model is used for augmentation as a result it suffers from the overfitting as well as vanishing gradient problem which makes the slower training process \cite{mascarenhas2021comparison, li2020classification}.
     \item $E_l$ described in Eq. \ref{eq7} does not clear the representation of style by the Gram matrix \cite{li2017demystifying}.
     \item $L_{MMDstyle}$ described in Eq. \ref{eq9} does not address style diversity in the images.
    \item The traditional DL models are black-box models and can not give enough context for the model's specifics \cite{sarp2023xai}.
     \item Existing NST-based augmentation methods need high-performance computational resources and time and the effort involved in choosing a 
    style image \cite{oza2022image}.
    \item  It is difficult to compare new methodologies because there are no augmentation techniques that explain the model's specifics.
\end{itemize}
\section{Material and Methods}
\label{sec:Material and Methods}
\begin{figure}[t]
\centerline{\includegraphics[width=0.5\textwidth]{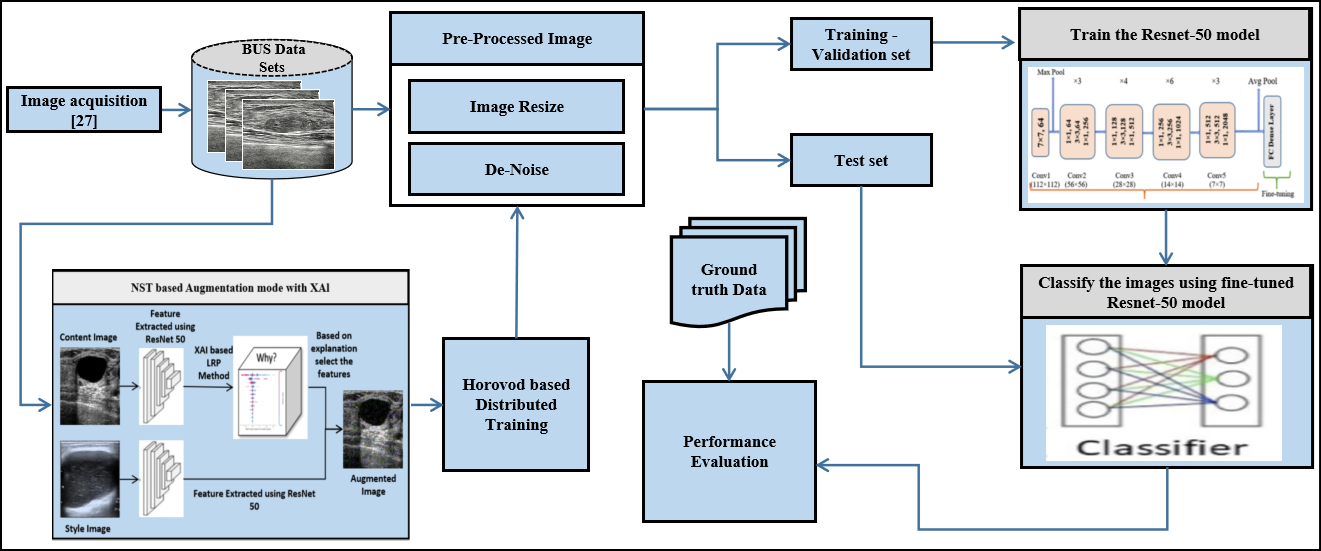}}
\caption{Workflow for the proposed model.}
\label{fig1}
\end{figure}

\begin{figure}[t]
\centerline{\includegraphics[width=0.5\textwidth]{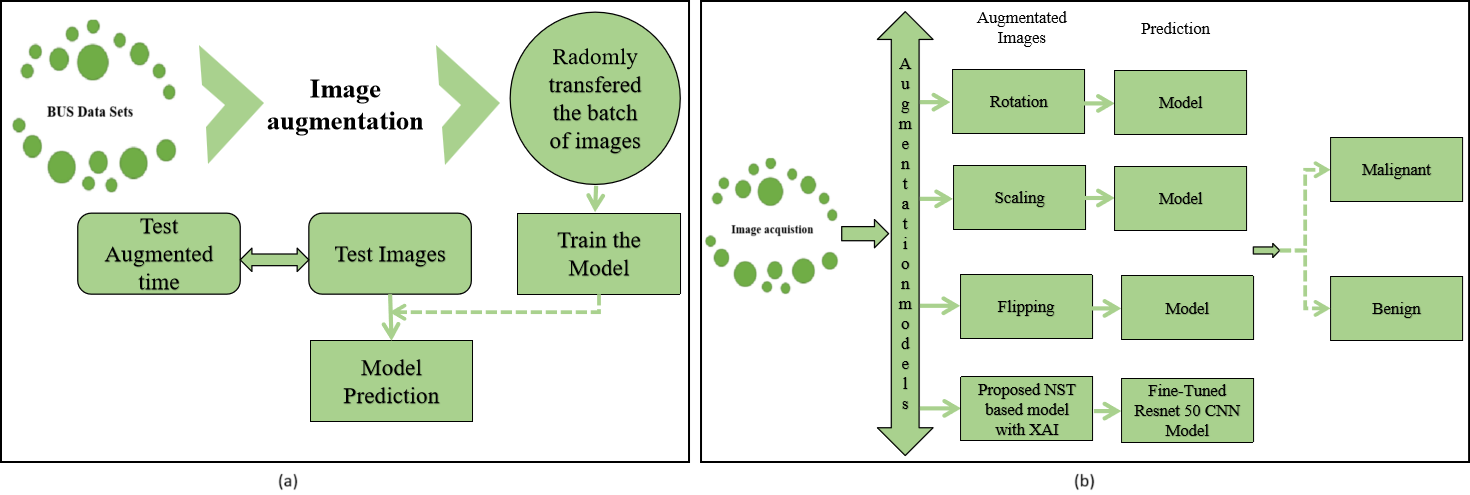}}
\caption{Training and testing of proposed image augmentation model.}
\label{fig2}
\end{figure}

This section provides a summary of the resources and techniques utilized in the different stages of BUS image augmentation and classification. The diagram illustrating the framework of the proposed model can be found in Fig. \ref{fig1}. Further, the augmentation block described in Fig. \ref{fig1}, is briefly presented in Fig. \ref{fig2}, which demonstrated how the proposed augmentation model was trained and tested. Initially, we combining the style loss from demystifying NST (DNST) and $mr^{2}$ NST model with ResNet50 and reformulate the style loss by using the second-order degree polynomial kernel for augmenting the images. Further, to explain the contributions of each input feature from the content images we utilized the LRP technique. This technique calculates relevance scores by utilizing the activation of each layer. These scores serve to highlight the significance of individual pixels or features within the input image toward generating the final output of the model. In the next step, we leveraged the Horovod framework to distribute the training of the proposed augmentation model, enabling scaling across an 8-GPU DGX cluster. This approach resulted in a significant acceleration, achieving a remarkable 5.09 speedup compared to single-GPU training. Further, we employed the SRAD filter in augmented BUS images to reduce the speckle noise. For each input image, this filter produces an illustration with several scales. As the scale progresses, the noise is systematically reduced, which enhances image clarity and quality. Next, we classify the denoised augmented images with a fine-tuned ResNet50 model to improve the classification accuracy. Finally, we compare our proposed XAI-based augmentation model with the existing cutting-edge augmentation methods.

\subsection{Data demographics and preprocessing}
\label{sec:Data demographics and preprocessing}
The proposed model is evaluated on a publicly available database comprising 800 BUS images, of which 348 are benign and 452 are malignant. Among these images, 167 are sourced from a public repository 
The second dataset, obtained from \cite{al2020dataset}, consists of 633 images.

The imaging instruments employed for scanning breast lesions are the LOGIQ E9 and the LOGIQ E9 Agile ultrasound systems, renowned for their high-quality imaging capabilities in radiology, cardiology, and vascular care,  especially when it comes to the scanning of breast lesions. These devices, which have a resolution of 1280 * 1024 pixels, provide clear images that are necessary for precise diagnosis. They make use of the ML6-15-D matrix linear probe's 1–5 MHz transducer.

\subsubsection{Bus image filtering}
\label{sec:Bus image filtering}
In medical imaging, particularly in BUS images, noise is a persistent issue that compromises the quality of extracted features. These images often suffer from speckle noise, leading to blurred edges, poor contrast, and reduced resolution, thereby preventing the effectiveness of CAD systems. To address this challenge, denoising is a crucial step. In this research, we employ the SRAD filter \cite{panigrahi2024mbccf, panigrahi2019ultrasound, panigrahi2022evaluation} to diminish noise and speckle across at different levels, producing a multi-scale image series. The selection of SRAD is deliberate because it effectively reduces noise while preserving texture information by halting diffusion across sharp edges as well as making an illustration in several scales for every image such that the noise consistently decreases with increasing scale. This approach aims to enhance the diagnostic accuracy and overall performance of CAD systems in analyzing BUS images. 
\subsection{Proposed augmentation model}
\label{sec:Proposed augmentation model}

\subsubsection{Proposed style loss function}
\label{sec:Proposed style loss function}
In this proposed model, by enlarging Eq. \ref{eq4}, we obtained Eq. \ref{eq16}. Further, we reconstructed the style loss function given in Eq. \ref{eq3}, by combining $L_{MMDstyle}$ in Eq. \ref{eq9} and $L_{multi-ref}$ in Eq. \ref{eq6} and \ref{eq7}. The proposed $L_{ps}$ presented in Eq. \ref{eq13}.
\begin{equation}
\label{eq13}
L_{ps}=\dfrac{1}{4N_l^2}MMD^2\sum_{ij}([F_l,F_l(I_s)])^2_{ij}
\end{equation}

\subsubsection{LRP propagation in ResNet50 model}
\label{sec:LRP propagation in Resnet50 model}

Let the neurons of the CNN, represented by the Eq. \ref{eq14}.
\begin{equation}
\label{eq14}
a_{i,j,k}=\sigma{(\sum_{m=1}^{M}\sum_{n=1}^{N}\sum_{c=1}^C w_{m,n,c,k}a_{i+m-1,j+n-1,c}+b_k)}
\end{equation}

where, the activation of neuron $a_{i,j,k}$ is represented at the position \textit{(i,j)}. Position in the $k^{th}$ feature map, \textit{m} and \textit{n} represent the height and width of the kernel respectively, $a_{i+m-1,j+n-1,c}$ neuron activation of the previous layer, $b_k$ and $w_{m,n,c,k}$ are the bias and weight of the neuron, and $\sigma$ is the activation function.

\begin{figure}[t]
\centerline{\includegraphics[width=0.5\textwidth]{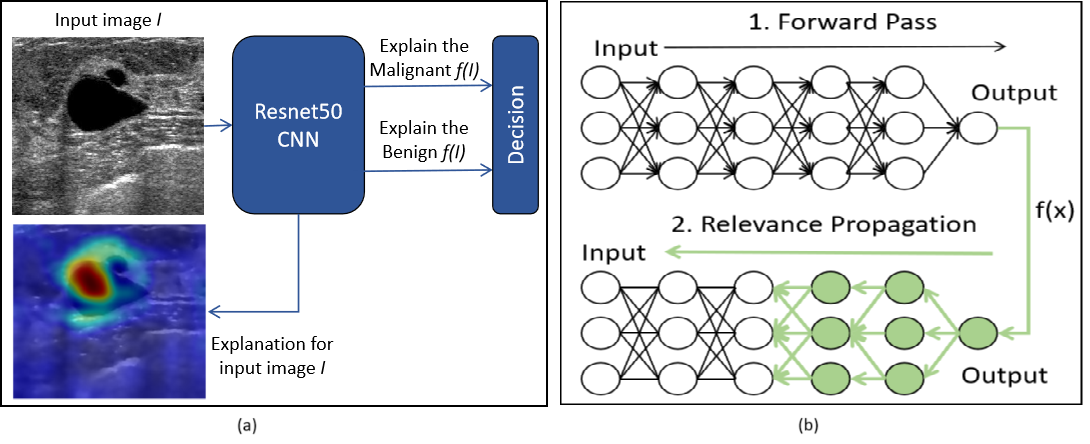}}
\caption{(a) Explanation of the prediction \textit{f(I)} for a given image \textit{I} and (b) architecture of LRP method.}
\label{fig3}
\end{figure}

LRP aims to assign a relevance score $R_p$ to each pixel \textit{p} in a given content image of \textit{x} in conjunction with a predictor function \textit{f} such that $f(x)\approx \sum_p R_p$. The $R_p>0$ represents the pixel, positively contributes to the output and $R_p<0$ indicates the negative contributes to the output. The pictorial representation of this score is presented as an image named a heatmap and the architecture of the LRP method is demonstrated in Fig. \ref{fig3} (a) and Fig. \ref{fig3} (b) respectively.

Additionally, the Eq. \ref{eq14} constituted with $\alpha$-$\beta$ rule for formulating the messages $R_{i\leftarrow j}^{(l,l+1)}$ (discussed in Sec \ref{sec:Layer-wise relevance propagation (LRP)}) shown in Eq. \ref{eq15}.

\begin{equation}
\begin{split}
\label{eq15}
R_{i+M-1,j+n-1,c}^{(l,l+1)}=\\
\sum_{i,j,k}\biggl(\alpha \dfrac{a_{i+m-1,j+n-1,c}w_{m,n,c,k}^+}{\sum_{m^{\prime},n^{\prime},c^{\prime}}a_{i+m^\prime-1,j+n^\prime-1,c^\prime}w_{m^\prime,n^\prime,c^\prime,k}^+}-\\
\beta \dfrac{a_{i+m-1,j+n-1,c}w_{m,n,c,k}^-}{\sum_{m^{\prime},n^{\prime},c^{\prime}}a_{i+m^\prime-1,j+n^\prime-1,c^\prime}w_{m^\prime,n^\prime,c^\prime,k}^-}\biggl)R_{i,j,k}^{(l)}
\end{split}
\end{equation}
where, $()^+$ and $()^-$ presents the positive and negative score respectively.

\subsubsection{Distributed augmented model training using Horovod Frame work}
\label{sec:Distributed augmented model training using Horovod Frame work:}
The training of our augmentation model is computationally intensive, necessitating a focus on distributed training to enhance scalability. To this end, we employed synchronized data parallelism to facilitate the scaling of the augmentation model training across multiple GPUs.

To make our single-GPU implementation to a Horovod-based multi-GPU distributed training framework, we pursued the following steps:

We began by initializing Horovod with the function hvd.init(), subsequently assigning a GPU to each of the TensorFlow processes. Then we encapsulated the TensorFlow optimizer within the Horovod optimizer using the command opt = hvd.DistributedOptimizer(opt). This Horovod optimizer is responsible for managing gradient averaging through a ring-based all-reduce mechanism. Finally, we synchronized the initial variable states by broadcasting them from the rank 0 process to all other processes using the hvd.callbacks.BroadcastGlobalVariablesCallback(0) method.
These steps ensured that our model training is efficiently distributed across multiple GPUs, thereby improving scalability and performance.

\subsection{Classification and performance evaluation of augmented images}
\label{sec:Classification of augmented images}
A model's performance can be improved while requiring less intensive training and careful data annotation when it is fine-tuned. It is in the classification of medical images \cite{davila2024comparison} where the images are usually complex, sparse such as BUS images. According to the classification results presented in \cite{davila2024comparison} the fine-tuned ResNet50 model has certain advantages over other benchmark techniques. Therefore, a fine-tuned ResNet50 \cite{davila2024comparison, hossain2022transfer} is employed in this research for classifying augmented images. However, the existing fine-tuned ResNet50 model is trained with different image datasets such as ImageNet and other medical images. Thus, in this study to classify the pre and post augmented images, we employed a ResNet50 architecture. Initially, we split the image data set into train, validation, and test sets. Then, trained the ResNet50 model with the BUS images (discussed in \ref{sec:Data demographics and preprocessing}). For breast cancer classification, we applied full fine-tuning since pretrained features from ImageNet are not optimal, and the distribution shift is minimal. By tuning all layers with the new data, the model can adapt and learn specific features (e.g., textures, patterns, abnormalities) from breast images, generally resulting in improved performance metrics (accuracy, precision, recall). Further, for the performance evaluation we used Eqs. (8)-(13) from \cite{panigrahi2024mbccf} to assess the performance of the classification model using different metrics.
\section{Experimental results and discussion}
\label{sec:Experimental results and discussion}
\subsection{Experimental setup and Evaluation Metrics}

\begin{figure}[t]
\centerline{\includegraphics[width=0.5\textwidth]{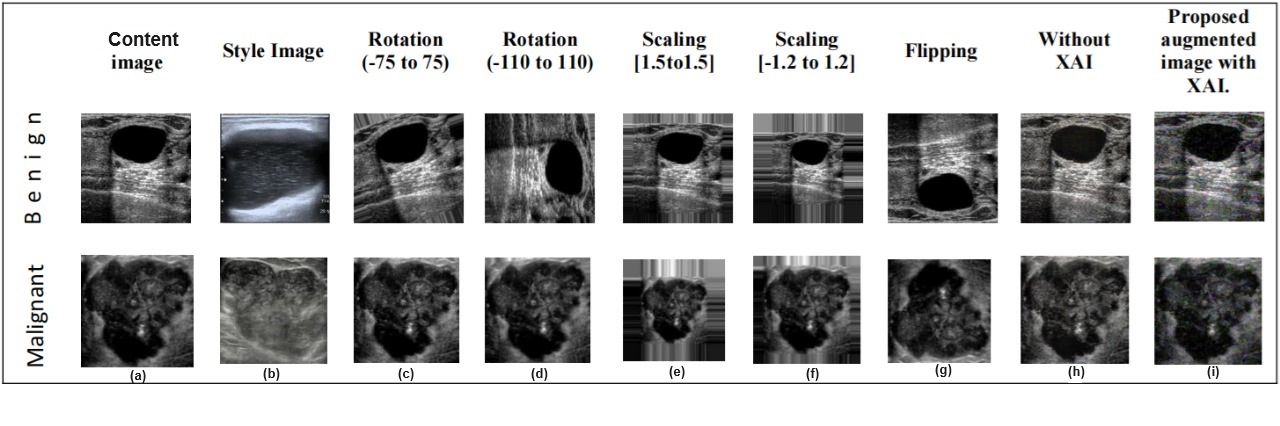}}
\caption{Implementation of different geometric and advanced image augmentation techniques.}
\label{fig4}
\end{figure}

\begin{figure}[t]
\centerline{\includegraphics[width=0.5\textwidth]{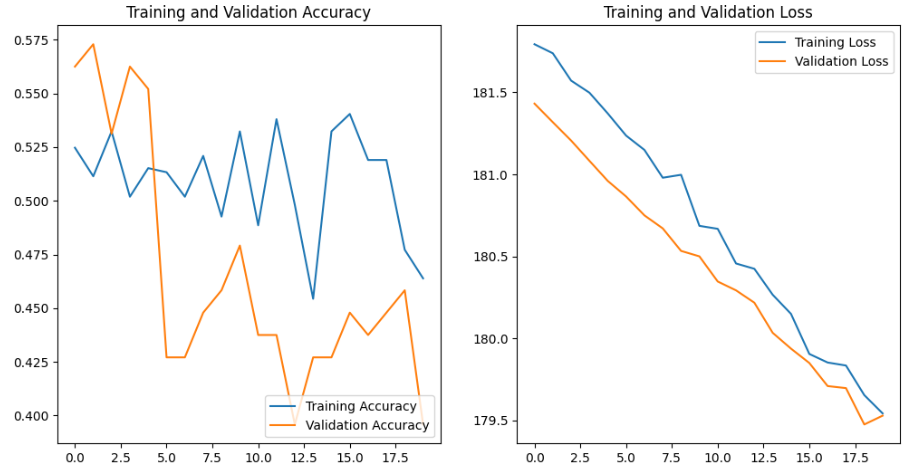}}
\caption{Classification model performance before augmentation.}
\label{fig9}
\end{figure}

\begin{figure}[t]
\centerline{\includegraphics[width=0.5\textwidth]{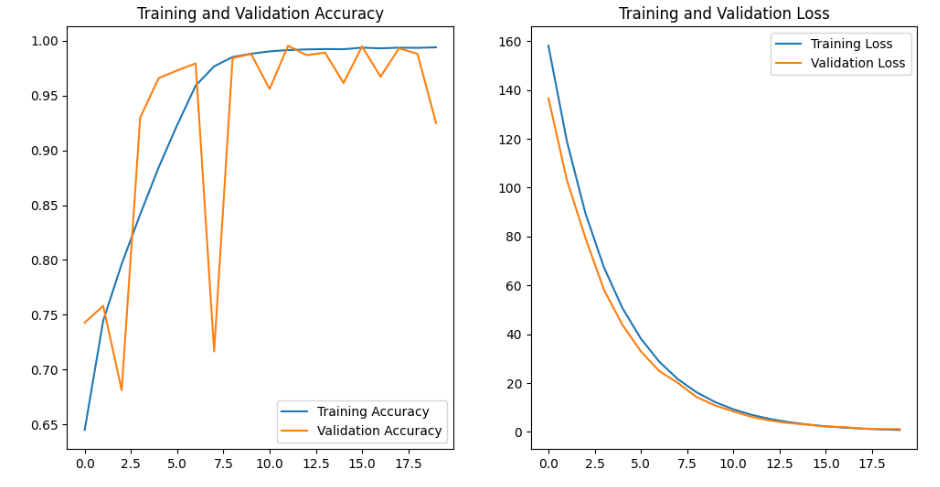}}
\caption{Classification model performance after proposed XAI-based augmentation.}
\label{fig10}
\end{figure}

\begin{table}[t]     
\begin{tabular}{p{2cm}p{1cm}p{1.1cm}p{1.2cm}p{1.5cm}}
\hline\noalign{\smallskip}
Aug model& sub category&Pre-aug Dataset Size&Post-aug Dataset Size&Post-aug model performance \\\hline
\centering \multirow{3}{*}{\parbox{2.3cm}{\raggedright Geometric Transform \cite{kalaivani2023geometric}}}\\
&Rotation&800 & 4000& Improved  by 13\\
&Scaling&800 & 4000& Increased by 15\\
&Flipping&800 &4000 &  Raised  by 12\\
NST with Vgg19 \cite{kavitha2021neural}& -&800 &800 & Improved  by 13 \\
$mr^{2}$NST \cite{wang2020mr}&-&800&1600&Improved  by 17\\
Proposed NST method without XAI&- &800 &325,408 & Enhanced by 20 \\
Proposed NST method with XAI (LRP)&- &800 &325,408 & Enhanced by 37.26 \\
\hline
\end{tabular}
\caption{Pre and post-augmentation dataset scaling impact on model accuracy (\%).}
\label{table4}  
\end{table}
\begin{table}[t]
\centering   
\begin{tabular}{|p{1.7cm}|p{1.5cm}|p{1.5cm}|p{1.3cm}|}
\hline\noalign{\smallskip}
\centering\multirow{2}{*}{No of process}& Parallel Time (Sec) &Sequential Time (Sec)&Speedup \\\cline{2-4}
 &$T_p$&$T_s$&S=$T_p/T_s$\\\hline
 
1& 3771.53 &\multirow{5}{*}{3771.53} &1\\
2&1953.60& &1.93\\
4&1166.73& &3.23\\
6&847.10& &4.45\\
8&739.65& &5.09\\
\hline
\end{tabular}
\caption{Multiprocessor based Augmentation model.}
\label{table5}    
\end{table}

\renewcommand{\arraystretch}{1.6} 
\begin{table}[t]
\centering
\begin{tabular}{p{2cm}p{0.9cm}p{0.6cm}p{0.6cm}p{0.7cm}p{0.7cm}p{0.7cm}}
\hline\noalign{\smallskip}
Augmentation method& sub category&Acc &Recall&  Specif-icity &Preci-sion &F1-score\\\hline
\centering \multirow{3}{*}{\parbox{2.3cm}{\raggedright Geometric Transform \cite{kalaivani2023geometric}}}
&Rotation&83.06 &82.01 & 82.33& 83.56&82.78 \\ 
&Scaling&85.48 &84.22 &84.11 &85.43 &84.79 \\
&Flipping&81.45 &80.27 &80.67 &81.55 & 80.86\\
NST with Vgg19 \cite{kavitha2021neural}& - & 83.34 & 82.70 & 81.33 & 83.11 & 82.92\\
$mr^{2}$NST \cite{wang2020mr}&-&89.23&87.45&87.78&88.61&88.08\\
Proposed NST method without XAI&- &90.71& 91.34&92.45 &90.45&90.84  \\
Proposed NST method with XAI (LRP)&- &92.47 &94.11 & 99.88& 92.50& 93.20\\
\hline
\end{tabular}
\caption{Quantitative evaluations for augmentation algorithms (\%)}
\label{table3}   
\end{table}

\begin{figure}[t]
\centerline{\includegraphics[width=0.5\textwidth]{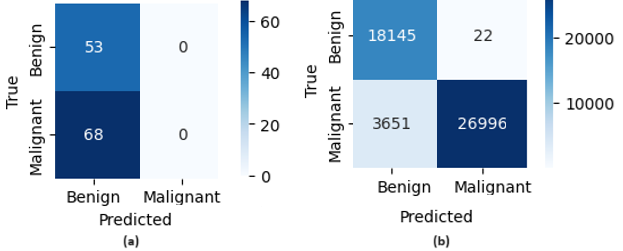}}
\caption{(a) and (b) are the confusion matrix of the classification model before and after augmentation respectively.}
\label{fig11}
\end{figure}

This section presents a detailed description of the experimental setup used and performance evaluation of the proposed augmentation model along with the classification of augmented data for BUS images. All experiments were executed utilizing Python, MATLAB R2024a, and Dev C++, operating on a Dell laptop furnished with a $13^{th}$ Generation Intel(R) Core(TM) i9-13900HX processor clocked at 2.20 GHz, paired with 16 GB of RAM. Further, to speed up the model an NVIDIA DGX A100 machine is equipped with dual CPUs, each hosting four A100 GPUs is used.

This study utilized two variations of the ResNet50 model: the first is a pre-trained ResNet50 used for augmenting the BUS images, and the second is a fully fine-tuned ResNet50 utilized to classify the datasets, where, the dataset were split into 70\% training, 15\% validation, and 15\% test set for both before and after augmentation as shown in Fig,\ref{fig1}. 

We collected a dataset comprising 800 BUS images. Further, to illustrate the optimal balance between minimizing noise and retaining texture details, we applied the SRAD filter with parameters set to an iteration interval of $n_t=10$ and a number of scales $L=8$. The implementation of the SRAD filter was carried out using MATLAB and DEV C++. Following the denoising process, a fully fine-tuned ResNet50 classifier is employed to classify the BUS images. Since the existing ResNet50 classifier model was trained with ImageNet datasets, thus we trained the ResNet50 model with the BUS images and found a classification accuracy of 55.21\%. The lower accuracy is due to overfitting. To resolve this issue, we implemented a novel XAI-based NST augmentation model on the original image. 

\subsection{XAI-based NST augmented model training
using Horovod framework}
This augmentation process unfolds in three distinct phases, outlined as follows:

\textbf{Step 1:} The style loss function of the NST model is designed by combining $L_{MMDstyle}$ and $L_{multi-ref}$. This is not only clear the representation of style by the Gram matrix but also addresses style diversity in the images. In the proposed augmentation model the weight $ w_l$ is set as 1.0 and the no of iteration is set as 1000, the loss function $\alpha$ and $\beta$ in Eq. \ref{eq1} is set as 1 (since the content losses from various approaches are same) and $\gamma \beta^ \prime$ (to maintain the balance between content and style matching) respectively and used ResNet50 in place of Vgg19 for extract the features from content and style images. 

\textbf{Step 2:} Propagate the LRP method in the ResNet50 model to explain the contributions of each input feature from the content
images. 

\textbf{Step 3:} To address the computational intensity of the XAI-based augmentation method we have utilized a distributed training approach leveraging Horovod that evenly split the content images among the GPUs where each process works on its subsets. For this, we have applied an NVIDIA DGX A100 machine equipped with dual CPUs, each hosting four A100 GPUs.

 To prove the efficacy of the augmentation model we compare our model with other cutting-edge methods shown in Fig. \ref{fig4}. Figs. \ref{fig4} (a) and (b) show the content and style image samples of benign and malignant images respectively. Figs. \ref{fig4} (c)-(g) represents the results of geometric augmentation methods like rotation, scaling, and flipping respectively. Figs. \ref{fig4} (h) and (i) show the augmented results by a proposed NST model without XAI and with XAI. 

To determine the efficient improvement in the augmentation model Table \ref{table4} presents the pre and post augmentation dataset size and model performance in terms of accuracy. It is observed that the specified augmentation model is not only able to increase the data size from 800 to 325,408 but also enhance the accuracy by 37.26\% as a result the model solved the overfitting issue as well as the vanishing gradient problem which makes faster the training process.

\textbf{Horovod framework:} To enhance the computational efficiency of the augmentation model, parallelization is employed. This made it possible for the process to be carried out more quickly. We observed a notable acceleration of up to 5.09 times in a parallel computing environment when compared to a sequential operation. We were able to generate 325,408 images in a total of 739.65 seconds with 8 GPUs. Table \ref{table5} displays the acceleration in performance and the duration of parallel processing for this task, respectively.

\subsection{ Quantitative evaluation of ResNet50 model for augmented image classification}
In this section, we present the numerical findings derived from the proposed augmentation model. Additionally, we performed a comparative analysis between the outcomes of our presented approach and those of currently available augmentation techniques.  The associated quantitative results are highlighted in Table \ref{table3}, showcasing various performance metrics including accuracy (Acc), recall, specificity, precision, and F1-Score.

A fully fine-tuned ResNet50 (trained by the collected BUS dataset) is used to classify the augmented images and evaluate the model performance. Using the proposed augmentation approach the data size was expanded to 325,408 images (121,104 benign and 204,304 malignant). Further, these augmented images are following the same pre-processing steps as before augmentation, including SRAD filtering, and were then split into training, testing, and validation sets, and classified using fully fine-tuned ResNet50. To achieve optimal performance and improved classification accuracy, the proposed model was trained for 20 epochs using early stopping after 7 consecutive epochs without validation loss improvement. We utilized the LeakyRelu activation function (due to accepting negative input), SGD optimizer with an initial $\eta$ = 0.0001 and momentum = 0.9, which subsequently reduced by \textit{ReduceLROnPlateau} function to enhance convergence. To ensure consistency, compatibility, and future scalability, \textit{categorical cross-entropy} with one-hot encoding is employed. A batch size of 32 is chosen to balance GPU memory efficiency and parallelization. To prevent overfitting and optimize training, we incorporated dropout (0.5) and batch normalization. A \textit{GlobalAveragePooling2D} is used to connect the layers, followed by a final dense layer with 2 units and softmax activation for classification. 

Fig. \ref{fig9} and \ref{fig10} present classification model performance before and after augmentation respectively. Fig. \ref{fig9} shows that before augmentation the model struggles with learning, showing instability in accuracy and only minimal decreases in loss, suggesting that augmentation plays a critical role in improving model performance. However, Fig. \ref{fig10} shows significant improvements, with high and stable accuracy and rapidly decreasing loss, indicating effective learning and good generalization. Further, before augmentation, the model archives a training and validation accuracy of 46.39\% and 39.58\% whereas, after augmentation training and validation accuracy improved to 99.38\%, and 92.49\% respectively and the test dataset resulted in a test accuracy of 92.47\%.

We examine the confusion matrix of benign, and malignant for the ResNet50 classification models shown in Fig. \ref{fig11}. Fig. \ref{fig11} (a) shows that before augmentation 53 samples were correctly classified as benign and 68 samples were incorrectly classified as malignant rest of the part are 0. Whereas, Fig. \ref{fig11} (b) shows a more balanced performance after augmentation, with a significant number of correct predictions for both benign and malignant cases.

 The performance of several augmentation models in terms of ACC, recall, specificity, precision, and F1-score is presented in Table \ref{table3}. It can be noticed that this study obtained higher accuracy due to the $L_{ps}$ being applied along with LRP. The aim of using LRP is the conservation principle and ability to identify the negatively relevant. This technique calculates relevance scores by utilizing the activation of each layer. These scores serve to highlight the significance of individual pixels or features within the input image towards generating the final output. 
\section{Conclusion}
\label{sec:Conclusion}
In this paper, we have introduced an innovative augmentation model which automatically augments the BUS images. This paper addresses the limitations of demystifying NST (DNST) and $mr^{2}$ NST
method by the proposed objective function and gives the context for the model's specifics.  Furthermore, it enhances efficiency by leveraging Horovod across multiple GPUs, significantly reducing computational time. We demonstrate that, when compared with other standard augmentation techniques, the suggested model outperforms better and is more precise. The results of our experiments demonstrate the superior accuracy and effectiveness of the proposed model over existing benchmark augmentation algorithms. 

This study can be enhanced in the future in a number of ways, including (1) Evaluation on larger datasets to strengthen confidence in performance, (2) Soliciting expert input to further enhance results, (3) Exploring the application of our technique to 3-D BUS images for improved the classification accuracy and performance of the CAD system.

\bibliography{aaai25}

\clearpage

\appendix
\section{Rephrasing the style loss function}
\label{sec:Rephrasing the style loss function}
\begin{equation}
\begin{split}
\label{eq16}
L_{s}=\dfrac{1}{4{N_{l}^2}M_{l}^2}\sum_{i=1}^{N_l}\sum_{j=1}^{N_l}{\biggl(\sum_{k=1}^{M_l} F_{l_ik}(I)^TF_{l_jk}(I)-\sum_{k=1}^{M_l} G_{l_{ik}}(I_{s_{ik}})G_{l_{jk}}(I_{s_{jk}})\biggl)}^2\\
=\dfrac{1}{4{N_{l}^2}M_{l}^2}\sum_{i=1}^{N_l}\sum_{j=1}^{N_l}\biggl({\biggl(\sum_{k=1}^{M_l} F_{l_ik}(I)^TF_{l_jk}(I)\biggl)}^2+{\biggl(\sum_{k=1}^{M_l} G_{l_{ik}}(I_{s_{ik}})G_{l_{jk}}(I_{s_{jk}})\biggl)}^2-\\
2\biggl(\sum_{k=1}^{M_l} F_{l_ik}(I)^TF_{l_jk}(I)-\sum_{k=1}^{M_l} G_{l_{ik}}(I_{s_{ik}})G_{l_{jk}}(I_{s_{jk}})\biggl)\biggl)\\
=\dfrac{1}{4{N_{l}^2}M_{l}^2}\sum_{i=1}^{N_l}\sum_{j=1}^{N_l}\sum_{k_{1}=1}^{M_l}\sum_{k_{2}=1}^{M_l}(F_{l_ik_{1}}(I)^TF_{l_jk_{1}}(I) F_{l_ik_{2}}(I)^TF_{l_jk_{2}}(I)+\\
G_{l_{ik_{1}}}(I_{s_{ik_{1}}})G_{l_{jk_{1}}}(I_{s_{jk_{1}}})G_{l_{ik_{2}}}(I_{s_{ik_{2}}})G_{l_{jk_{2}}}(I_{s_{jk_{2}}})-\\
2F_{l_ik_{1}}(I)^TF_{l_jk_{1}}(I)G_{l_{ik_{2}}}(I_{s_{ik_{2}}})G_{l_{jk_{2}}}(I_{s_{jk_{2}}})\\
=\dfrac{1}{4{N_{l}^2}M_{l}^2}\sum_{k_{1}=1}^{M_l}\sum_{k_{2}=1}^{M_l}\sum_{i=1}^{N_l}\sum_{j=1}^{N_l}(F_{l_ik_{1}}(I)^TF_{l_jk_{1}}(I) F_{l_ik_{2}}(I)^TF_{l_jk_{2}}(I)+\\
G_{l_{ik_{1}}}(I_{s_{ik_{1}}})G_{l_{jk_{1}}}(I_{s_{jk_{1}}})G_{l_{ik_{2}}}(I_{s_{ik_{2}}})G_{l_{jk_{2}}}(I_{s_{jk_{2}}})-\\
2F_{l_ik_{1}}(I)^TF_{l_jk_{1}}(I)G_{l_{ik_{2}}}(I_{s_{ik_{2}}})G_{l_{jk_{2}}}(I_{s_{jk_{2}}})\\
=\dfrac{1}{4{N_{l}^2}M_{l}^2}\sum_{k_{1}=1}^{M_l}\sum_{k_{2}=1}^{M_l}\biggl( (\sum_{i=1}^{N_l}F_{l_ik_{1}}(I)^TF_{l_ik_{2}}(I)^T)^2+(\sum_{i=1}^{N_l}G_{l_{ik_{1}}}(I_{s_{ik_{1}}})G_{l_{ik_{2}}}(I_{s_{ik_{2}}}))^2-\\
2(F_{l_ik_{1}}(I)^TG_{l_{ik_{2}}}(I_{s_{ik_{2}}}))^2\biggl)\\
=\dfrac{1}{4{N_{l}^2}M_{l}^2}\sum_{k_{1}=1}^{M_l}\sum_{k_{2}=1}^{M_l}\biggl((F_{lk_{1}}(I)^TF_{lk_{2}}(I))^2+(G_{l_{k_{1}}}(I_{s_{k_{1}}})^TG_{l_{k_{2}}}(I_{s_{k_{2}}}))^2-2(F_{lk_{1}}(I)^TG_{l_{k_{2}}}(I_{s_{k_{2}}}))^2\biggl)
\end{split}
\end{equation}

\end{document}